\documentclass{article}

\usepackage[nonatbib,final]{neurips_2019}

\usepackage[utf8]{inputenc} % allow utf-8 input
\usepackage[T1]{fontenc}    % use 8-bit T1 fonts
\usepackage{hyperref}       % hyperlinks
\usepackage{url}            % simple URL typesetting
\usepackage{booktabs}       % professional-quality tables
\usepackage{amsfonts}       % blackboard math symbols
\usepackage{nicefrac}       % compact symbols for 1/2, etc.
\usepackage{microtype}      % microtypography
\usepackage{graphicx, booktabs}
\usepackage{amsmath,amssymb}
\usepackage{multirow}
\usepackage{subfigure}
\usepackage{hhline}
\usepackage{xcolor}
\usepackage{colortbl}
\usepackage{caption}
\title{Pi-PE: A Pipeline for Pulmonary Embolism Detection using Sparsely Annotated 3D CT Images}
\author{%
 Deepta Rajan, David Beymer, Shafiqul Abedin and Ehsan Dehghan \\
 IBM Research AI, \\
 Almaden, San Jose, CA 95120. \\
Email: \texttt{\{drajan, beymer, sabedin, edehgha\}@us.ibm.com}
}

\begin{document}

\maketitle

\begin{abstract}
Pulmonary embolisms (PE) are known to be one of the leading causes for cardiac-related mortality. Due to inherent variabilities in how PE manifests and the cumbersome nature of manual diagnosis, there is growing interest in leveraging AI tools for detecting PE. In this paper, we build a two-stage detection pipeline that is accurate, computationally efficient, robust to variations in PE types and kernels used for CT reconstruction, and most importantly, does not require dense annotations. Given the challenges in acquiring expert annotations in large-scale datasets, our approach produces state-of-the-art results with very sparse emboli contours (at 10mm slice spacing), while using models with significantly lower number of parameters. We achieve AUC scores of $0.94$ on the validation set and $0.85$ on the test set of highly severe PEs. Using a large, real-world dataset characterized by complex PE types and patients from multiple hospitals, we present an elaborate empirical study and provide guidelines for designing highly generalizable pipelines.
\end{abstract}

\section{Introduction}
\label{sec:intro}

A pulmonary embolism (PE) manifests as blocks in pulmonary arteries triggered by blood clots, air bubbles, or accumulation of fat tissues that occur typically during surgery, pregnancy or cancer. PE is known to be one of the leading causes of cardiac-related mortality, where an early diagnosis and treatment is expected to have a significant impact in controlling the mortality rate. It is estimated that between $300,000$ to $600,000$ individuals are affected by PE every year in the US \cite{beckman2010venous}. Computed tomographic pulmonary angiography (CTPA) is the primary diagnostic exam to detect arterial diseases, given the high spatial resolution of CT scanners. Each CTPA study is a $3$D image containing hundreds of slices, some of which show evidence of PE as irregularly shaped filling defects.

%dark pixel regions partially surrounded by a border of lighter pixels. 

In practice, each occurrence of PE can belong to one of the following broad categories: peripheral, segmental, subsegmental, lobar, or saddle type, which can be typically determined based on its arterial location. In particular, subsegmental PEs are considered to be the hardest to detect, since they often occur subtly in subsegmental branches of the pulmonary artery. Consequently, radiologists are required to painstakingly examine every slice in a CT image for detecting PEs, thus making this process highly cumbersome and time-consuming. Moreover, unlike other common diseases visualized in chest CTs such as lung nodules, which usually appear spherical, or emphysema, which can be observed across the entire lung, PEs are known to appear much more asymmetrically only in isolated regions of pulmonary vasculature. 

Given the afore-mentioned challenges in detecting PEs, computer-aided diagnostic tools~\cite{liang2007computer} have become prominent. More specifically, data-driven approaches based on deep learning have produced promising results for automatic PE detection~\cite{tajbakhsh2015computer}. The most successful solutions in medical image analysis often comprise multi-stage pipelines tailored for a specific organ/disease. Without loss of generality, such a pipeline in turn includes a segmentation stage for candidate generation, i.e. semantically meaningful regions that are likely to correspond to the disease occurrence, and a classification stage for the actual detection. A crucial bottleneck of this approach is the need for large annotated datasets. Acquiring expert annotations for $3$D volumes, where every instance of disease occurrence is annotated (often referred to as \textit{dense annotations}), is time-consuming and error-prone. Furthermore, there is a dearth of standard benchmark datasets for PE detection using CTPA, and most of the research in this space is conducted on custom datasets or small-scale challenge dataset \cite{cadpe}. In practice, the small-data challenge is often combated by adopting a transfer learning strategy that refined classifiers pre-trained on natural image data~\cite{tajbakhsh2016convolutional}. Recently, Huang \textit{et al.} \cite{penet} showed that such a strategy, where $500$K video clips were used to pre-train a $77$-layer $3$D convolutional network, could be effectively fine-tuned for learning a PE classifier using dense annotations.

\begin{figure}[t]
	\centering
	\includegraphics[width=0.77\textwidth]{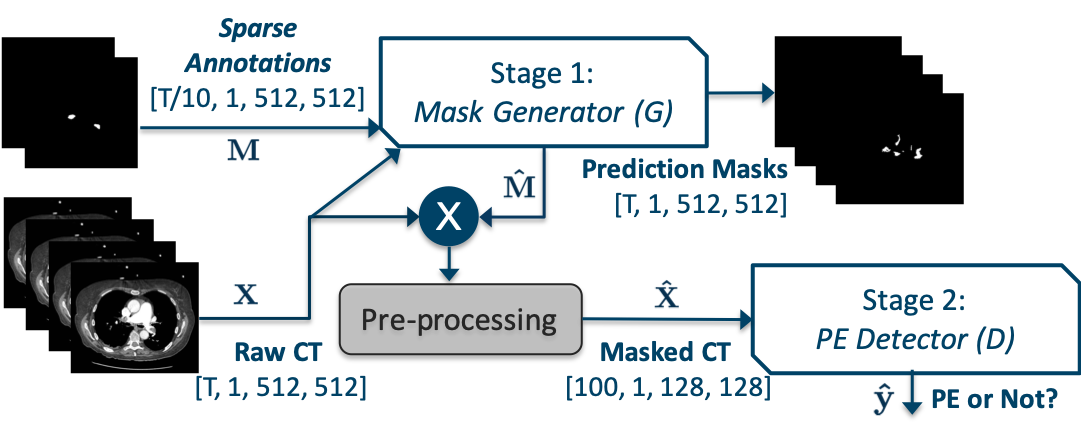}
	\caption{Illustration of the proposed two-stage pipeline for PE detection. \texttt{Stage 1} is comprised of a mask generator $G$, based on $2$D context-augmented UNets, and \texttt{Stage 2} is comprised of a PE detector $D$, involving a $2$D Conv-LSTM model coupled with multiple instance learning. Both the raw input CTs ($X$) and sparsely annotated masks $M$ are fed into $G$ to produce prediction masks ($\hat{M}$). Subsequently, $X$ and $\hat{M}$ are multiplied to obtain $\hat{X} = X \odot \hat{M}$ that is fed as input to \texttt{Stage 2}. Finally, $D$ outputs a  prediction $\hat{y} \in [1,0]$ indicating the presence or absence of PE.}
	\label{fig:pipeline}
\end{figure}

In this paper, we adopt an alternate approach of building an accurate PE detection system using only sparsely labeled CT volumes. More specifically, we develop a two stage detection pipeline (shown in Figure \ref{fig:pipeline}) designed exclusively using $2$D CNNs, wherein the candidate generation state utilizes a novel context-augmented U-Net and the classifier stage employs a simple $2$D Conv-LSTM model coupled with multiple instance learning~\cite{zhu2017deep, ilse2018attention, braman2018disease}, compared to the $77-$layer $3$D CNN in~\cite{penet}. We find that, even with significantly smaller number of parameters and with no pre-training, our approach produces state-of-the-art detection results on a challenging, large-scale real-world dataset. Further, we study its generalization across hospitals/datasets, given the large disparity across image acquisition systems and protocols, and demonstrate the proposed approach to be highly robust.

%\textbf{Contributions:} 
%In this paper, we design a two-stage approach to accurately detect all types of PE under a weak-supervision setting by leveraging both sparse instance-level semantic label as well as volume-level diagnostic label together in an end-to-end learning pipeline. Compared to existing approaches, our pipeline is made up of a simplified $20$-layer network while achieving state-of-the-art performance on a large, first-of-its-kind CTPA dataset containing studies acquired using multiple imaging protocols across various hospitals. In addition, our approach allows leveraging unannotated CT studies to further improve performance and model generalizability. 
Our contributions can thus be summarized as follows:
\begin{itemize}
\item We develop a novel two-stage approach for PE detection -- \texttt{Stage 1} is comprised of a $2$D UNet based \cite{ronneberger2015u} mask generator and \texttt{Stage 2} utilizes a \textit{ConvLSTM} \cite{xingjian2015convolutional} based PE detector.
\item Our approach does not require expensive \textit{dense annotations} and operates exclusively on \textit{sparse annotations} generated for every 10 mm of positive CT scans.
\item We use a context-augmentation strategy that enables the $2$D U-Net in \texttt{Stage 1} to produce high-quality masks.
\item By modeling each $3$D CT volume as a bag of instances, i.e., features for each $2$D slice obtained using a Conv-LSTM, we propose to employ multiple instance learning, based on feature aggregation, to detect PE. 
\item For the first time, we evaluate our approach using a large-scale, multi-hospital chest CT dataset that well represents real-world scenarios through the inclusion of complex PE types and diverse imaging protocols.
\item We present insights from an elaborate empirical study, while discussing the impact of different architectural design choices on the generalization performance.
\item We show that our approach achieves state-of-the-art detection performance, with AUC scores of $0.94$ on a validation set of all PE types and $0.85$ on a test set of high-severity PE.
\end{itemize}

% MIL decision fusion is a problem and people suggest doing feature fusion
% Also better to build an end-to-end system..
% Not using unlabeled samples is an issue…
% Need to leverage instance-level semantic label and volume-level diagnostic label together in an end-to-end learning pipeline

% \textbf{Contributions:} In this paper, we propose an efficient two-stage pipeline for accurately detecting all types of PE, while full-supervision is not assumed. Compared to existing approaches, our pipeline is made up of a simplified architecture while achieving state-of-the-art performance on a large, first-of-its-kind CTPA dataset containing studies acquired using multiple imaging protocols across various hospitals. \texttt{Stage$1$} of the pipeline comprises of a $2$D UNet mask generator ($G$) and \texttt{Stage$2$} includes a multiple instance learning (MIL) based PE detector ($D$). The detector uses a convolutional-LSTM \cite{xingjian2015convolutional} architecture to model each $3$D CT study as a bag of instances, while features from every instance is aggregated using a feature aggregator based on the \textit{max} function. Our contributions can be summarized as follows:

% \item We build an efficient two-stage deep model while maintaining reduced complexity through use of fewer layers to enable faster training at lower annotation needs.

% , while features from every instance is aggregated using a feature aggregator based on the \textit{max} function.

\section{Dataset Description}
\label{sec:data}
\begin{figure}[t]
	\centering
	\includegraphics[width=0.75\linewidth]{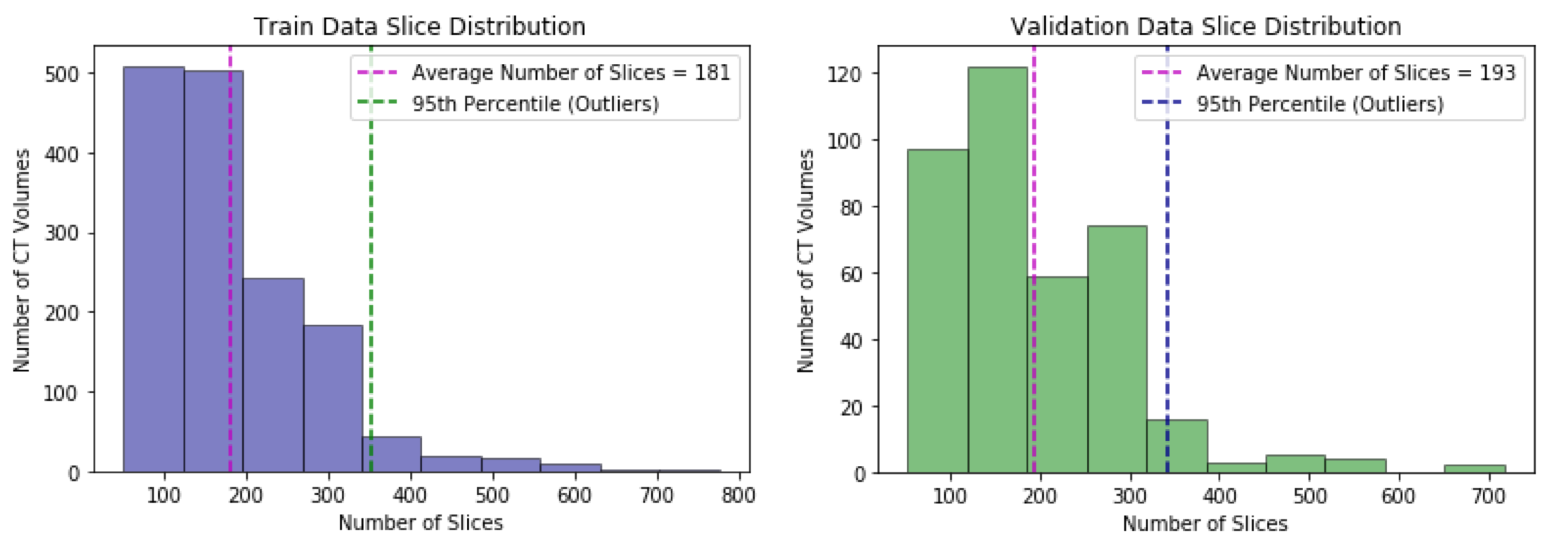}
	\caption{Histogram of number of image slices in each CT study across the train and validation datasets. Our PE detector ($D$) as part of \texttt{Stage 2} is designed to use CTs of $100$ slices during training.}
	\label{fig:slice_dist}
\end{figure}
\begin{table}[t]
	\centering
	\renewcommand{\arraystretch}{1.3}
	\begin{tabular}{l|c|c|c|c}
		\hline
		\multicolumn{1}{c|}{\cellcolor{gray!15}\textbf{Label}} &
		\cellcolor{gray!15}\textbf{Train} & \cellcolor{gray!15}\textbf{Validation} & \cellcolor{gray!15}\textbf{Test} &
		\cellcolor{gray!15}\textbf{Total} \\ \hline
		Positive                                & 1,053                  & 264                        & 385    & 1,702              \\ 
		Negative                           & 473                  & 118                        & 127        & 718     \\ \hline 
	\end{tabular}
	\caption{Sample sizes of the custom CT dataset sourced from multiple hospitals.}
	\label{table:data-size}
	\vspace{-5pt}
\end{table}
% \begin{itemize}
%     \item data collection and annotation process
%     \item protocols and what the dataset encompasses
%     \item data sizes, data issues, study selection
%     \item preprocessing
%     \item why is this study is the most representative among works reported so far?
%     \item annotations available only every 10 mm, 
%     % image values -1024 HU and 500 HU mapped to 0-255
%     \item protocols - contrast enhanced chest CT - PE protocol v/s CTA protocol, different dose levels (noise level), different image reconstruction kernels
%     \item DICOM kernel, filter, pixel spacing, slice thickness, axial view 
% \end{itemize}
We collected 1,874 PE positive and 718 negative anonymized, contrast-enhanced chest CT studies and their corresponding radiology reports. Note that, due to the specific anonymization protocol used in our data curation process, we are unable to determine if two studies belong to the same patient. Our dataset is curated to represent variations across multiple imaging centers ($>100$) and different contrast-enhanced imaging protocols, namely PE and CTA. In comparison, currently reported studies in the literature including the state-of-the-art PENet~\cite{penet} focus exclusively on the PE protocol, and study generalization to only two hospitals. Consequently, the problem setup we consider is significantly more challenging and further, we do not assume access to dense annotations. Table \ref{table:data-size} shows the sample sizes used for train, validation and test phases of our algorithm. Further, the number of slices in each of the volumes can vary significantly, as illustrated in Figure~\ref{fig:slice_dist}.

\subsection{Sparse Annotations}
As part of the data preparation, we adapt the NLP pipeline described in~\cite{guo2017efficient} to identify PE positive studies from a patient's radiology report, while detecting the radiologist recommended phrase \textit{'no evidence of pulmonary embolism'} to identify PE negative studies. The positive studies were further vetted by board-certified radiologists who annotated the scans by drawing a contour around every embolism occurrence on slices approximately 10mm apart. This process naturally results in multiple unannotated slices between every pair of annotated slices, depending on the slice spacing. We refer such CT studies to be \textit{sparsely} annotated. While each study was annotated by only one clinical expert, a total of $17$ radiologists served as annotators in the process. Out of the 1,874 positive studies that were processed, $172$ of those were discarded due to reasons including the lack of definitive evidence for presence of PE (discrepancy between annotator and the reporting radiologist), insufficient contrast, metal or motion artifacts, etc.

%Hence, we cannot obtain an accurate estimate of the total number of unique patients. 
%As part of the data preparation, we adapt the NLP pipeline described in~\cite{guo2017efficient} to detect the presence of PE from a patient's radiology report. 

\section{Proposed Methodology}
\label{sec:method}
\begin{table}[t]
	\centering
	\renewcommand{\arraystretch}{1.2}
	\begin{tabular}{|c|c|c|c|c|}
		\hline
		\textbf{Subsegmental PE} &
		\textbf{Segmental PE} &
		\textbf{Lobar PE} &
		\textbf{Ground Truth} &
		\textbf{CA U-Net}\\
		\hline
		\includegraphics[width=0.15\linewidth]{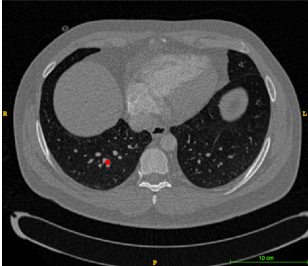}& \includegraphics[width=0.15\linewidth]{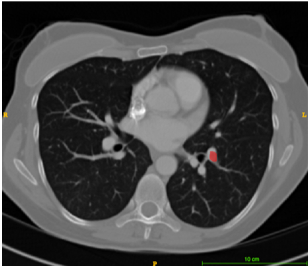} &
		\includegraphics[width=0.15\linewidth]{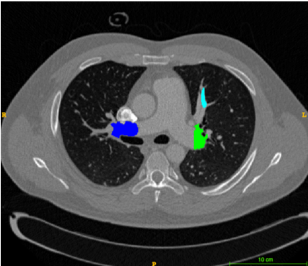} &
		\includegraphics[width=0.15\linewidth]{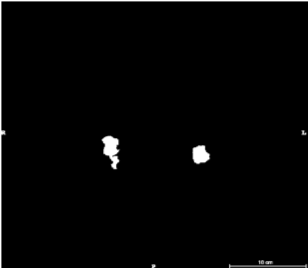} &
		\includegraphics[width=0.15\linewidth]{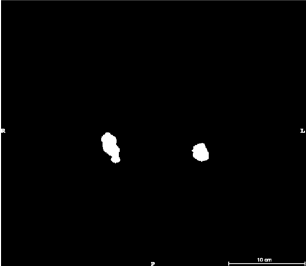} 
		\\ \hline
	\end{tabular}
	\caption{Example CT studies showing various types of PE. We also show the ground truth annotation and the prediction from \texttt{Stage 1} for the Lobar PE example.}
	\label{table:examples}
\end{table}

\subsection{Approach Overview}
We develop a two stage approach for PE detection from CT images. While the first stage processes the raw CT volumes to produce a mask that identifies candidate regions that are likely to correspond to emboli regions, the latter stage operates on the masked volume from \texttt{Stage 1} to perform the actual detection. In contrast to existing solutions, our approach relies exclusively on $2$D convolutions and does not require dense annotations. As illustrated in Figure~\ref{fig:pipeline}, \texttt{Stage 1} is implemented using a novel context-augmented $2$D U-Net, and for \texttt{Stage 2}, we adopt a multiple instance learning (MIL) formulation, wherein each $3$D volume $X$ is viewed as a bag of instances defined using the individual $2$D slices $x_1, \dots, x_T$. Here, $T$ denotes the total number of slices in $X$. Broadly, existing MIL methods focus on inferring appropriate aggregation functions either on (i) the instance-level predictions ($y_1, \dots, y_T$) to produce bag-level prediction $y$~\cite{zhu2017deep, braman2018disease}, or (ii) the instance-level latent features $\{z_1, \dots, z_T\}$ to construct the bag-level feature $z$, which can be subsequently used to obtain the prediction $y$~\cite{ilse2018attention}. We adopt the latter approach, where the instance features are obtained using a $2$D Conv-LSTM model and the feature aggregation is carried out using different functions including mean, max and learnable attention modules.

\subsection{Stage 1: Candidate Mask Generation}
The role of \texttt{Stage 1} is to segment an image and identify PE candidates which are localized regions with semantics indicative of the disease. As an initial preprocessing step, each input CT scan is resampled to a volume with $2$mm slice spacing. The architecture for the mask generator $G$ is a standard $2$D U-Net~\cite{ronneberger2015u}, an encoder-decoder style network comprised of a contracting path to downsample the input image while doubling the number of channels, followed by an expansive path to upsample the image. Though using a $2$D U-Net significantly simplifies the computation, processing each $2$D slice independently fails to leverage crucial context information in the neighboring slices. In order to circumvent this, we propose to extract \textit{slabs} of $4$ neighboring slices from either side of each $2$D slice, to form a stack of $9$ slices. We treat the raw intensities from each the $9$ slices as the channel dimensions, thus producing slabs of size ($9, 512, 512$) representing number of channels, height and width. We refer to this architecture as the context-augmented U-Net (CA U-Net). We observed from our experiments that this simple augmentation strategy consistently produced high-quality masks (see example in Table~\ref{table:examples}).

Each \textit{downblock} in our U-Net architecture contains $2$D convolution layers with a $3$x$3$ kernel, a batch normalization layer, a ReLU activation layer and a maxpool layer with a stride of $2$ to downsample the image. While, each \textit{upblock} upsamples and then concatenates features at the same level or depth of the network, followed by a convolutional layer coupled with batch normalization and ReLU activation. The depth of the network $G$ was fixed at $4$. Upon training, $G$ produces output probabilities for each pixel in the middle slice of the slab, indicating the likelihood of being PE candidate. The training objective was to achieve a high \textit{dice coefficient}, a metric which describes the pixel-wise similarity between prediction masks ($\hat{M}$) and ground truth annotation masks ($M$), and has a range of $[0-1]$. It is defined as $DC = (2*\sum_{i=1}^{N} \hat{m_i}m_i)/(\sum_{i=1}^{N} \hat{m_i}^{2} + \sum_{i=1}^{N} m_i^{2})$, where $N$ is number of voxels, $\hat{m_i} \in \hat{M}$ and $m_i \in M$ \cite{milletari2016v}. In practice, we adopt the continuous dice loss as $\mathcal{L}_{dice} = 1 - DC$.

%defined as $DC = 2*(|M \cap \hat{M}|)/(|M|+|\hat{M}|)$, a metric which describes the pixel-wise similarity between prediction masks ($\hat{M}$) and ground truth annotation masks ($M$), and has a range of $[0-1]$. In practice, we actually adopt the continuous dice loss as defined in \cite{milletari2016v}. $\mathcal{L}_{dice} = 1 - DC$.

% We find from Our \texttt{Stage$1$} model achieves a dice score of $0.66$ towards segmenting positive regions in an image.

% -- typically models plagued by false positives
% -- steps to reduce false positive
% -- what is detection rate?
% -- did we use dropout or augmentation?
% -- hyperparams, kernel, lr, number of steps

\subsection{Stage 2: Pulmonary Embolism Detection}
\begin{figure}[t]
	\centering
	\includegraphics[width=0.85\textwidth]{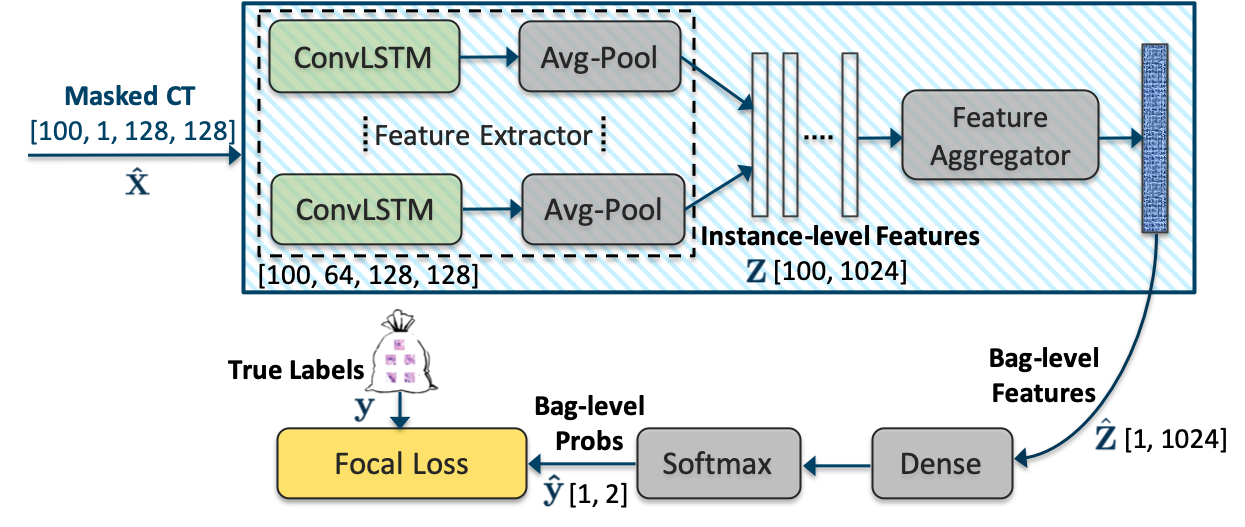}
	\caption{Architecture of the proposed \texttt{Stage 2} PE detector. Each instance ($\hat{x}_1, \dots, \hat{x}_{100}$) is transformed using a single \textit{ConvLSTM} layer followed by an \textit{AvgPool} layer to obtain instance-level features ($Z={z_1, \dots, z_T}$). A feature aggregation function (e.g. \textit{max}) then produces a bag-level feature that can be subsequently used for the actual classification.}
	\label{fig:pe-detector}
\end{figure}

As described earlier, to perform the actual PE detection, we treat each CT volume as a bag of multiple $2$D slices (instances). Hence, the goal of~\texttt{Stage 2} is to assign a prediction label to a bag indicating the presence or absence of PE. Multiple instance learning is a well-studied problem, where each instance is processed independently, and their features (or predictions) can be aggregated for obtaining bag-level predictions. However, we argue that processing each slice independently in a $3$D volume can produce noisy predictions since the local context is not included. More specifically, we utilize a Conv-LSTM~\cite{xingjian2015convolutional} model to produce instance features that automatically incorporates context from its neighboring slices, and perform feature aggregation similar to any MIL system.

As illustrated in Figure \ref{fig:pe-detector}, the PE detector $D$ contains an instance-level feature extractor followed by an MIL module. The feature extractor is a $2$D Conv-LSTM  architecture that effectively captures spatio-temporal correlations in a CT volume and produces meaningful instance-level features. All input-to-state and state-to-state transitions use a convolution operation containing $64$ filters, a $3$x$3$ kernel and a padding size of $1$. The input to $D$ are the masked CTs, denoted as $\hat{X}$, that is obtained as follows: First, the prediction masks, $\hat{M}$, from \texttt{Stage 1} are multiplied with raw CT volumes $X$ to create masked CT volumes. We then reduce the $z$-dimension of the masked volumes for computational efficiency. To this end, we use a lung segmentation algorithm to detect the boundary axial slices ($z_{start}, z_{end}$) that span the lung region. We then extract $T=100$ middle slices from within this range, crop to reduce image height and width to ($384,384$) and finally resize to ($128, 128$), thus transforming $X$ to produce $\hat{X} \in \mathbb{R}^{100,1,128,128}$. Each instance ($\hat{x}_1, \dots, \hat{x}_{100}$) is transformed by the \textit{Conv-LSTM} model as follows:
\begin{align}
\nonumber &\text{(input gate):\quad} i_{t} = \sigma(W_{xi}\hat{x}_{t} + W_{hi}h_{t - 1} + W_{ci}\circ c_{t - 1} + b_i) \\
\nonumber&\text{(forget gate):\quad} f_{t} = \sigma(W_{xf}\hat{x}_{t} + W_{hf}h_{t - 1} + W_{cf}\circ c_{t - 1} + b_f)
\\
&\text{(cell state):\quad} c_{t} = f_{t}\circ c_{t - 1} + i_{t}\circ\tanh{(W_{xc}\hat{x}_{t} + W_{hc}h_{t - 1}  + b_c)} 
\\
\nonumber&\text{(output gate):\quad} o_{t} = \sigma(W_{xo} \hat{x}_{t} + W_{ho} h_{t - 1} + W_{co}\circ c_{t} + b_o) 
\\
\nonumber&\text{(hidden state):\quad} h_{t} = o_{t} \circ \tanh{(c_{t})} \label{eqn:convlstm}
\end{align}The features are then average-pooled using a kernel of size $32$ to produce dense $1024$-dimensional features $z_1, \dots, z_{100}$ for all slices in $\hat{X}$. In order to perform feature aggregation for MIL, we explored the use of \textit{max}, \textit{mean} and learnable \textit{self-attention} functions. The self-attention function used is similar to the one described in~\cite{song2018attend}, and was implemented with multiple attention heads.
\begin{align}
\hat{z} = \sum_{k=1}^{100} a_k z_k; \quad a_k = \frac{\exp(U^T \tanh(V z_k^T))}{\sum_j \exp(U^T \tanh(V z_j^T))}.
\end{align}Here, $a_k$ denotes the attention coefficients and $U$, $V$ denote the learnable parameters for the attention module. The aggregated feature from the multi-head self-attention was further projected using a \textit{linear} layer to obtain the final bag-level features. For training the detector model $D$, we also explored using the standard binary cross-entropy (BCE) loss and the focal loss \cite{lin2017focal} defined as:
\begin{equation}
\ell(\hat{y},y) = -y(1-\hat{y})^{\gamma} \log(\hat{y}) \label{eqn:focalloss}
\end{equation}At inference time, we apply the preprocessing steps of cropping and resizing to $(128,128)$ spatial resolution, but make predictions for moving windows of $T=100$ slice (with $25$ slice overlap) and use the maximum detection probability as the final prediction for the test CT scan.

% hidden and cell states are four-dimensional denoted as $\hat{X}, h, c$

% stage $2$ includes a multiple instance learning (MIL) based classifier along with a feature aggregator. The classifier model uses a convolutional-LSTM architecture to model each $3$D CT study as a bag of instances, while features from every instance is aggregated using a self-attention mechanism.

% \begin{figure*}[t]
% 	\centering
% 	\subfigure[Loss function]{\includegraphics[width=0.45\linewidth]{figs/mit_loss.png}}
% 	\subfigure[Accuracy]{\includegraphics[width=0.45\linewidth]{figs/mit_acc.png}}
% 	\subfigure[Loss function]{\includegraphics[width=0.45\linewidth]{figs/ptb_loss.png}}
% 	\subfigure[Accuracy]{\includegraphics[width=0.45\linewidth]{figs/ptb_acc.png}}
% 	\caption{Training behavior -- Convergence characteristics of the proposed \texttt{DDxNet} model with the arrhythmia classification (top) and myocardial infarction detection (bottom) datasets.}
% 	\label{fig:training}
% \end{figure*}

\section{Empirical Results}
\label{sec:results}
In this section, we present a detailed empirical analysis conducted to evaluate the performance of the proposed pipeline and study its behavior with respect to different architectural choices. In particular, we share insights from ablation studies focused on effect of the number of instances $T$ used in \texttt{Stage 2} PE detection, the strategies used for feature extraction and aggregation, and finally the choice of loss function used for training \texttt{Stage 2}. 

\subsection{Experiment Setup}
All our experiments are based on modifying the PE detector in \texttt{Stage 2}, while retaining the \texttt{Stage 1} model to be the same. Details on sample sizes used in our empirical study are provided in Table \ref{table:data-size}. Typically, for successful adoption of detection algorithms in clinical practice, they are expected to have a high recall rate on the abnormal cases (also referred to as sensitivity). However, in order to obtain a well-rounded evaluation of the performance we report the following metrics: accuracy (\textit{Acc}), sensitivity or recall (\textit{Rec}) and precision (\textit{Prec}). %specificity (\textit{Spec})
$$
\text{Acc}=\frac{tp + tn}{tp + fp + fn + tn}; \quad \text{Rec} = \frac{tp}{tp + fn}; \quad \text{Prec} = \frac{tp}{tp + fp}; \quad \text{f1}=2. \frac{\text{Prec}. \text{Rec}}{\text{Prec} + \text{Rec}}, %\quad \text{Spec} = \frac{tn}{tn + fp},
$$where $tp, fp, fn, tn$ correspond to the number of true positives, false positives, false negatives and true negatives respectively. Further, to obtain an overview on the performance we use the $f1$-score and the area under receiver operator curve (AUROC). 
%Note, the $f1$-score can be measured as
%$$
%\text{f1}=2. \frac{\text{Prec}. \text{Rec}}{\text{Prec} + \text{Rec}}. %\quad \text{Prec} = \frac{tp}{tp + fp}.
%$$

\textbf{Training:} We trained the model $D$ for \texttt{Stage 2} using an adaptive learning rate of $1e-3$, which is subsequently reduced based on plateauing behavior of the validation loss. Other hyperparameters include a batch size of $8$, the number of instances ($T$) set to $100$ (unless specified otherwise), and the \textit{Adam} optimizer with a weight decay of $0.01$. All implementations were carried out in Pytorch, and we performed multi-gpu training using $4$ NVIDIA GTX GPUs.
\begin{table*}[t]
	\renewcommand{\arraystretch}{1.3}
	\renewcommand{\tabcolsep}{3.4pt} %12.5, 13.4pt
	\centering
	\begin{tabular}{l|c|c|c|c|c|c}
		\hline
		\multicolumn{1}{c|}{\cellcolor{gray!20}}        & \multicolumn{6}{c}{\cellcolor{gray!20} \textbf{PE Detector: Feature Extractor + Aggregation Strategy + Loss Function}} \\ \hhline{*{1}{>{\arrayrulecolor{gray!20}}-}>{\arrayrulecolor{black}}*{6}{-}}
		
		\multicolumn{1}{c|}{\multirow{-2}{*}{\cellcolor{gray!20} \textbf{Metrics}}} & {\cellcolor{blue!10} \textbf{C+SA+B}} & {\cellcolor{blue!10} \textbf{CL+SA+B}}      &  {\cellcolor{blue!10} \textbf{CL+MSA+B}} & {\cellcolor{blue!10} \textbf{CL+Mean+B}}         & {\cellcolor{blue!10} \textbf{CL+Max+B}} & {\cellcolor{blue!10} \textbf{CL+Max+F}}            \\ 
		\hline 
		\hline
		
% 		\multicolumn{6}{l}{\cellcolor{blue!10}\textbf{Task 1: Phenotyping}} \\
		
		\hline
% 		\small
		Accuracy  & 0.80  & 0.83   & 0.83   & 0.84   & 0.86  & \textbf{0.88} \\
		AUC       & 0.86  & 0.88   & 0.89    & 0.90  & 0.91  & \textbf{0.94} \\ 
		F1-Score & 0.86   & 0.88   & 0.87  & 0.88    & 0.90  & \textbf{0.91} \\ \hline
	\end{tabular}
	\caption{Validation performance comparison using different combinations of feature extractor, aggregation strategy and loss function. Here, CL = Conv-LSTM, C = Conv. with no LSTM, SA = Self-attention, MSA = Multi-head Self Attenion, B = BCE Loss, F = Focal Loss, Max = Max pooling aggregation and Mean = Average pooling aggregation.}
\vspace{-10pt}
\label{table:perf}
\end{table*}

\begin{figure*}[t]
	\centering
	\subfigure[Focal Loss]{\includegraphics[width=0.45\linewidth]{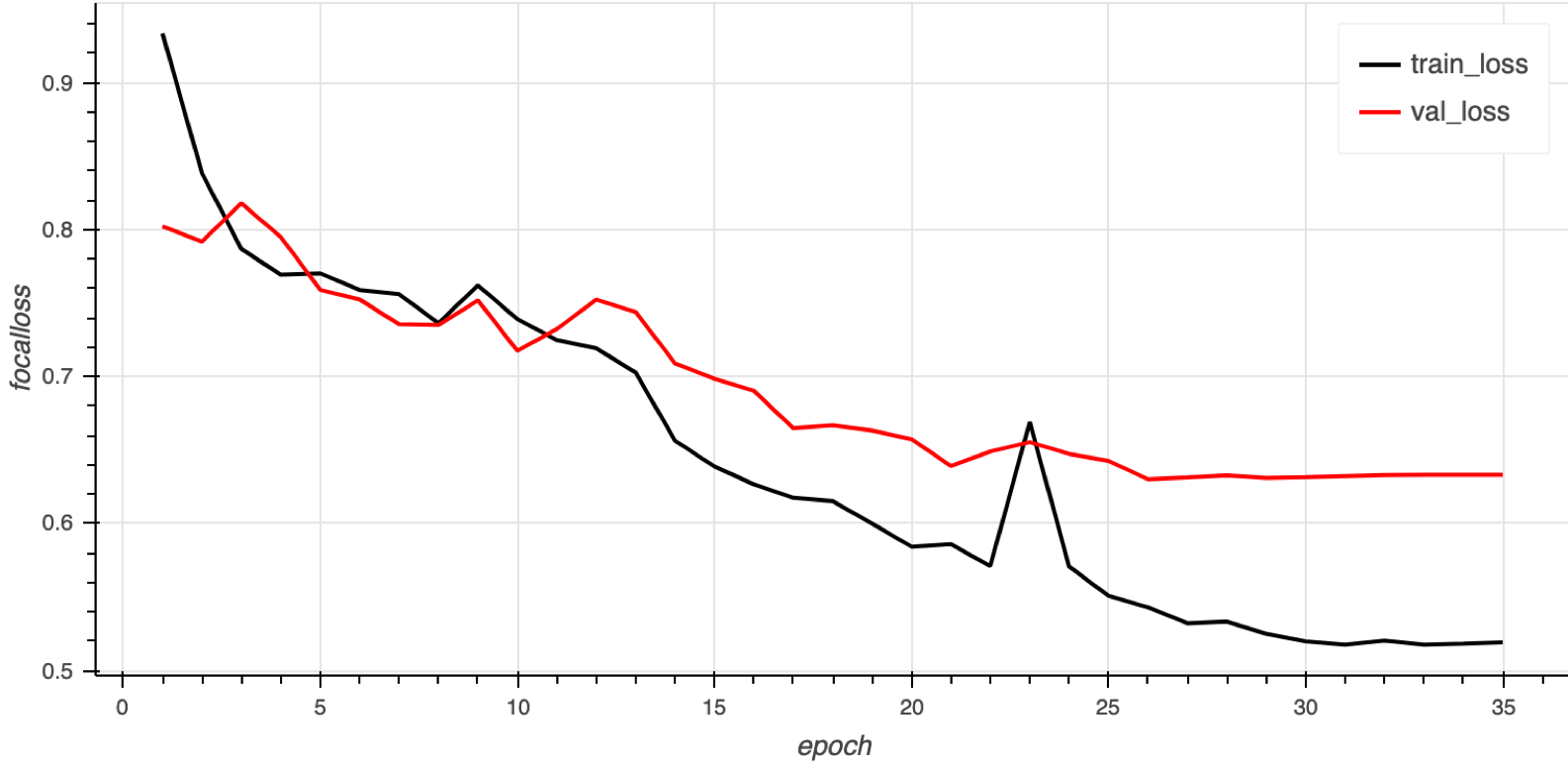}}
	\subfigure[AUROC]{\includegraphics[width=0.45\linewidth]{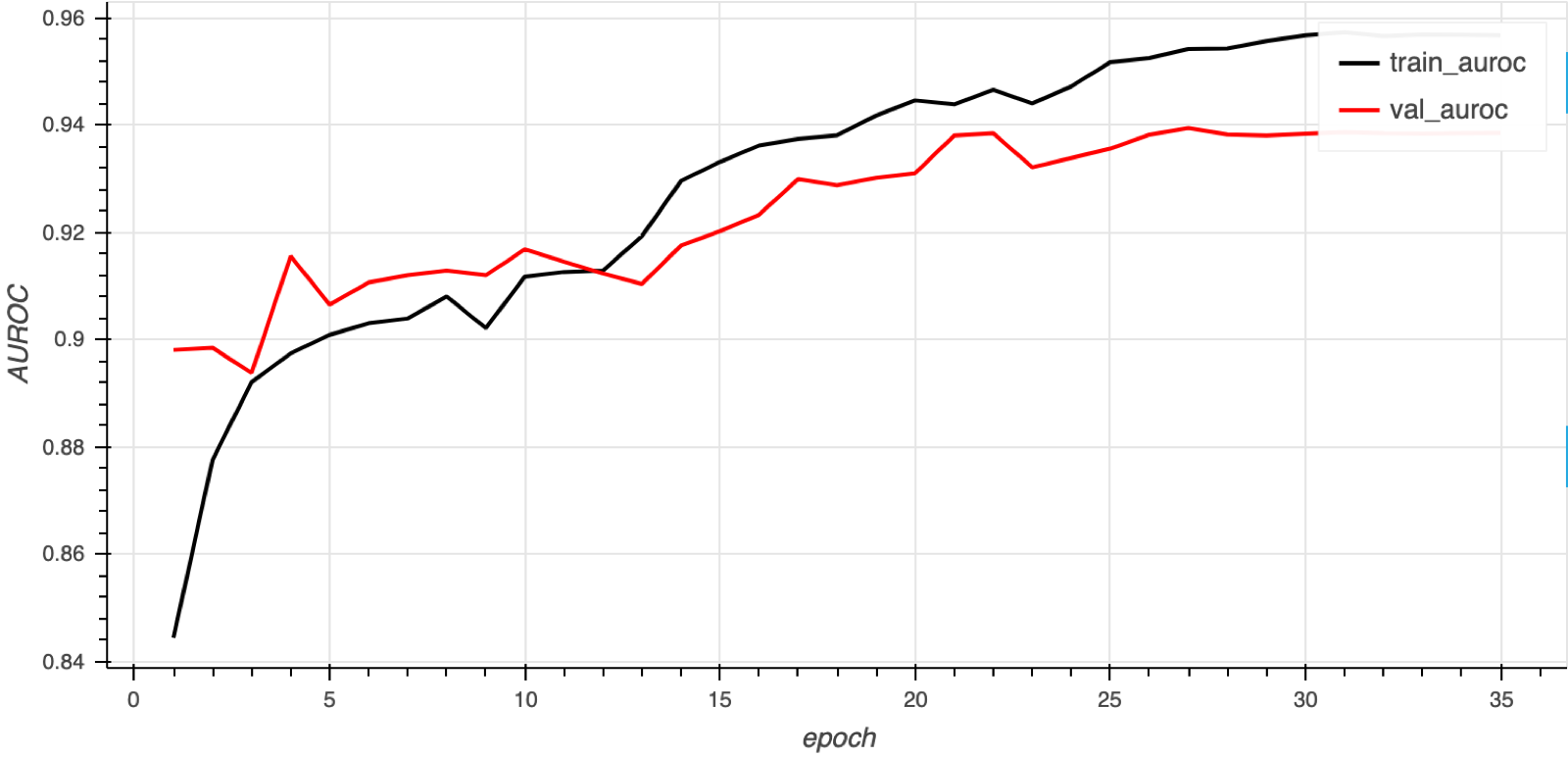}}
	\caption{Training behavior of the \texttt{Stage 2} PE detection model. Using the focal loss produces significantly better generalization when compared to the conventional BCE loss.}
	\label{fig:training}
\end{figure*}

% \begin{itemize}
%     \item Metrics: Accuracy, AUROC, AUPRC, Sensitivity, Specificity
%     \item Confusion matrices, ROC curves
%     \item GPUs, and rough train time
%     \item hyperparameters - $1e-3$ learning rate, batch size $8$, Adam optimizer, $64$ convolutional filters, $3$ kernel size
% \end{itemize}
\subsection{Ablation Studies}
\begin{figure*}[t]
	\centering
	\subfigure[Effect of $T$]{\includegraphics[width=0.45\linewidth]{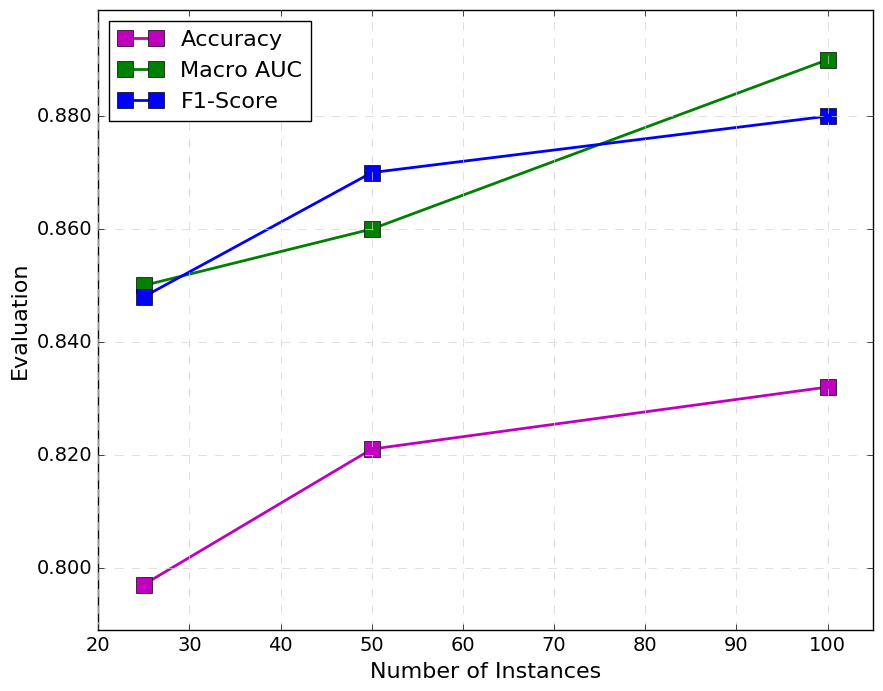}}
	\subfigure[Types of PE]{\includegraphics[width=0.45\linewidth]{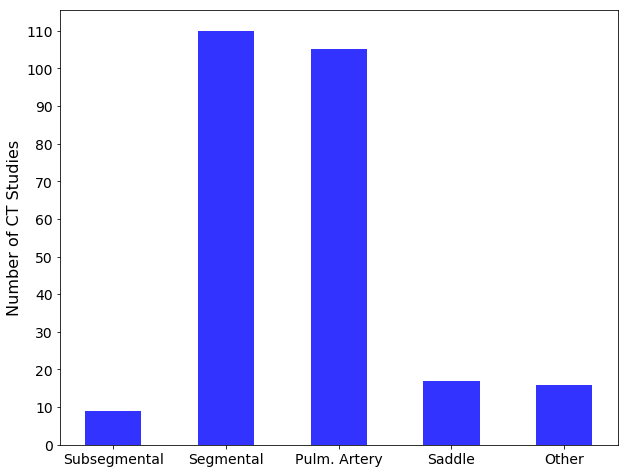}}
	\caption{Performance characteristics - (a) Increasing number of instances ($T$) from $25$ to $100$ steadily improves model performance, (b) Histogram of various types of PE.}
	\label{fig:inst}
\end{figure*}

In this section, we provide details on the various ablation studies carried out to understand the effect of each architectural component towards the validation performance of the PE detector. 
\begin{itemize}
\item \textbf{Study 1 - Effect of number of instances}: Given the limited GPU memory sizes, and the large sizes of CT volumes, we varied the number of instances ($T$) that were selected from the masked volume to invoke \texttt{Stage 2} and studied its effect on the performance. We found that increasing $T$ expectedly improved the classifier performance as shown in Figure \ref{fig:inst}(a).
\item \textbf{Study 2 - Feature Extraction and Aggregation}: We studied the effect of using LSTM for feature extraction by training a model with Conv-LSTM (CL) layer + Self-Attention (SA) and compared it to using only Conv. (C) + Self-Attention (SA). As expected, the Conv-LSTM model appears to extract more representative features from the slices, compared to treating each of the slices to be independent, as seen in Table \ref{table:perf}. A similar empirical analysis on the choice of feature aggregation strategy was carried out. Surprisingly, using \textit{max} pooling achieved the best performance when compared to even the self-attention module with learnable parameters. This is likely due to the fact that the LSTM already captures dependencies between instances in the bag, thus not requiring a dedicated attention module.
\item \textbf{Study 3 - Loss Functions}: We also observed that using Focal (F) loss, with $\gamma$ = $2$ in Equation \ref{eqn:focalloss}, significantly boosts the detection performance by countering the inherent imbalance in the dataset as opposed to using the conventional Binary Cross-Entropy (B) loss.
\end{itemize}

% \begin{itemize}
%     % \item Different choices for Stage-1: U-Net with slabs, U-Net without slabs, Pre-trained 2-D encoder
%     \item Choice for Stage-2: CLAM
%     \item Data imbalance study.. detector performance with different positive sample distribution: $25$\%, $50$\%, $75$\% and $100$\% positives --- finding is that using weighted random sampler, imbalance can be addressed. Another finding is that dataset inherently biased towards positives achieved better performance on $382$ validation set.
%     \item Trade off between timesteps, batch size and performance
%     \item Detector performance with and without Self-Attention, use Mean, Max pooling between timestep features like in Attention MIL paper
%     \item Detector performance with and without LSTM
% \end{itemize}

\begin{figure}[t]
	\centering
	\includegraphics[width=0.95\linewidth]{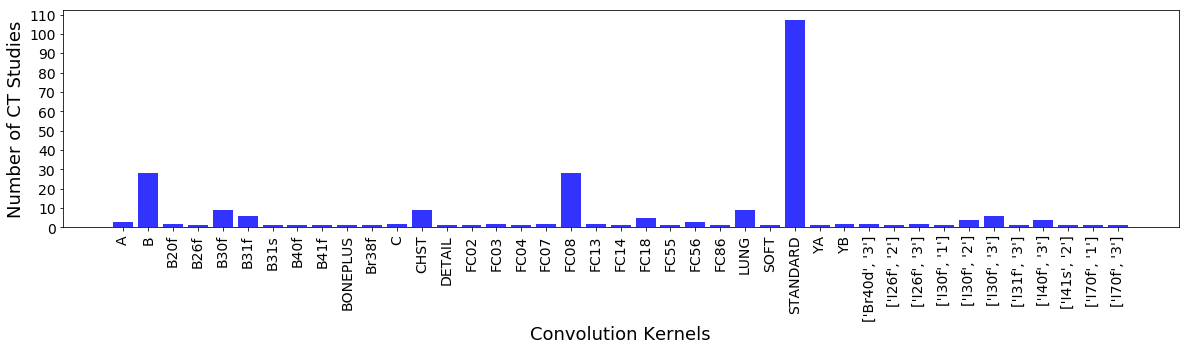}
	\caption{Test set distribution -- convolution kernels used for reconstructing the CT volumes.}
	\label{fig:hist}
\end{figure}

\subsection{Test Performance - Variations across PE Types}
Our dataset contains several kinds of PE with varying levels of severity, a distribution of which is shown in Figure \ref{fig:inst}(b). We report performance of our pipeline on low severity types such as subsegmental and segmental PE, as well as high severity types, namely saddle and main pulmonary artery shown in Figure \ref{fig:roc}(a). As expected, our pipeline picks up evidence for high severity PE more easily by achieving an AUC score of $0.85$, while obtaining an AUC of $0.70$ in detecting low severity PEs that are harder to find. When compared to the PENet model~\cite{penet}, our approach achieves improved test accuracies on a dataset characterized by larger amounts of variability, while using a significantly reduced number of parameters. Note that PENet has $28,398,705$ parameters~\cite{penet}, while our model only has $3,168,116$ parameters where \texttt{Stage 1} has $1,966,450$ and \texttt{Stage 2} has $1,201,666$.

%Note that PENet has $28,398,705$ parameters \cite{penet} while our model has only $1,966,450$ parameters in \texttt{Stage 1} and $1,201,666$ parameters in \texttt{Stage 2} with a total of $3,168,116$. 

%Note, number of parameters in PENet is $28,398,705$, while number of parameters in our pipeline is $3,168,116$ (\texttt{Stage 1}$=1,966,450$, \texttt{Stage 2} $=1,201,666$).

\subsection{Test Performance - Variations across CT Convolution Kernels}
\begin{figure*}[t]
	\centering
	\subfigure[AUROC (PE Types)]{\includegraphics[width=0.45\linewidth]{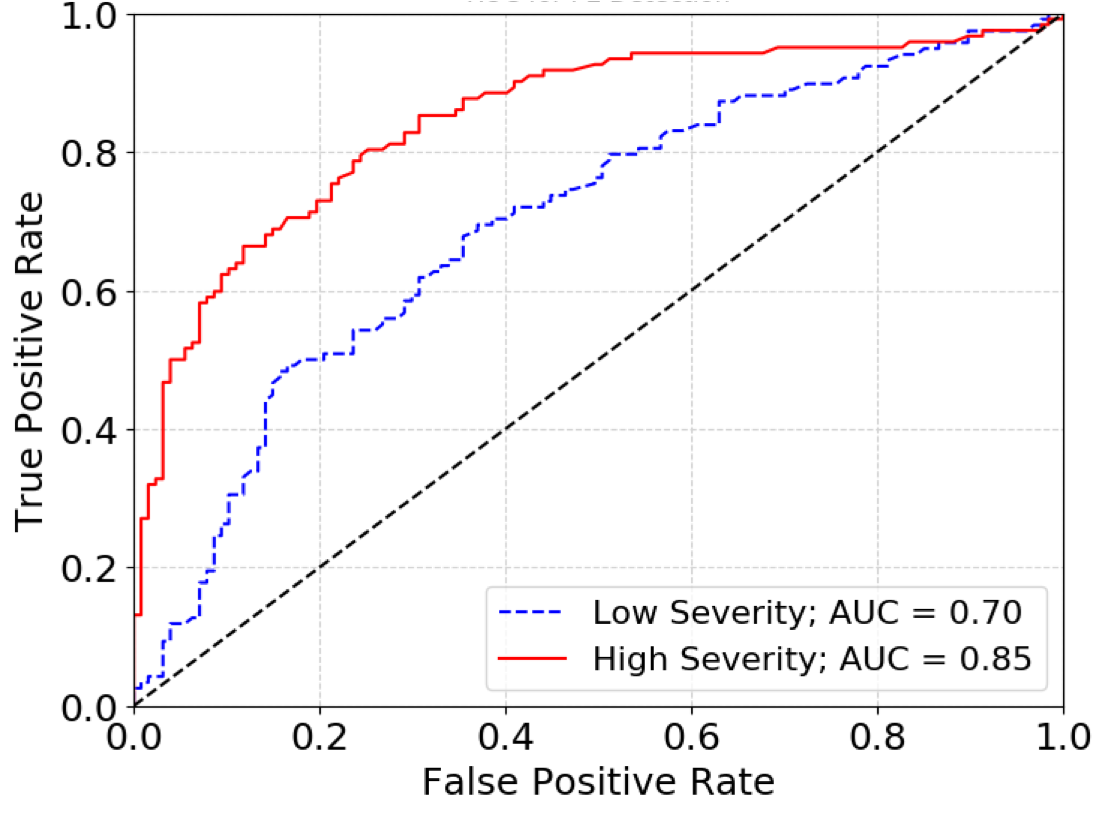}}
	\subfigure[AUROC (Kernel Types)]{\includegraphics[width=0.45\linewidth]{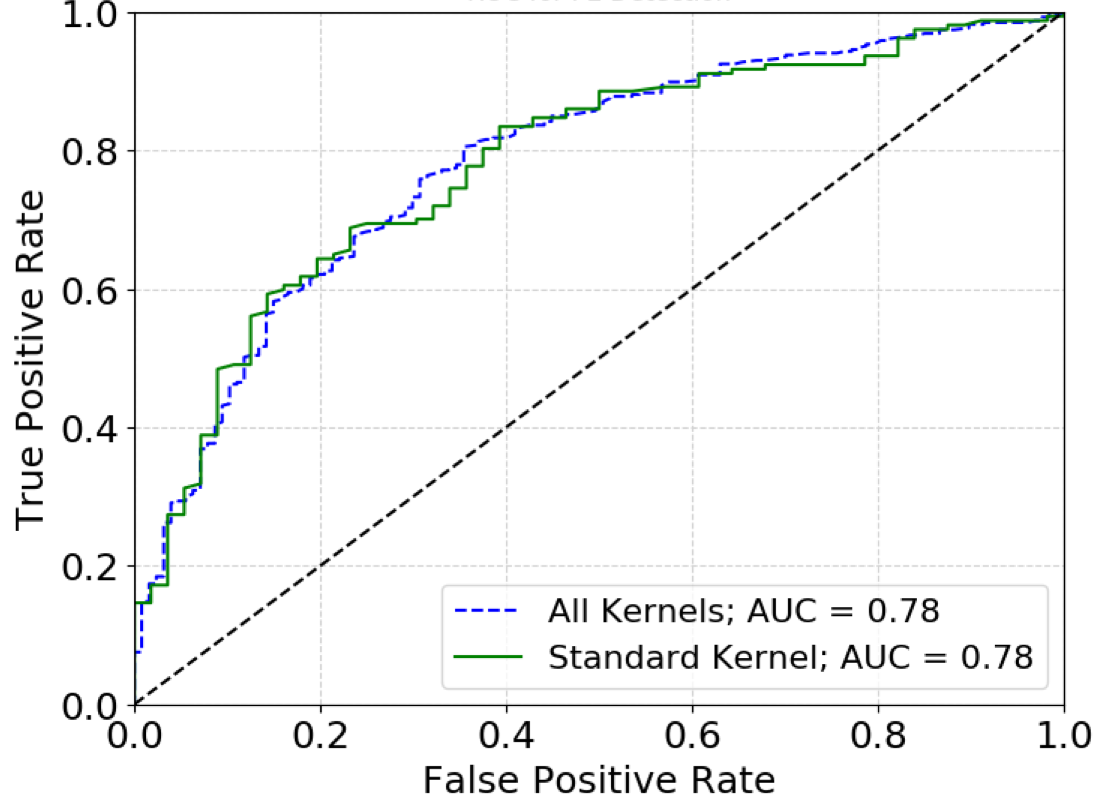}}
	\caption{Test set performance -- (a) AUROC when dataset includes all types of PE, versus only mild PE, (b) AUROC when dataset includes CT studies with diverse convolution kernels versus cases reconstructed using only the GE standard kernel.}
	\label{fig:roc}
\end{figure*}

In addition, our dataset is comprised of CT images reconstructed using different convolutional kernels, whose choice typically controls the image resolution and noise-levels. Figure \ref{fig:hist} shows the distribution of kernels for our dataset, where despite most cases using the 'GE Standard' kernel, the dataset includes volumes reconstructed using a wide variety of other kernels. From Figure \ref{fig:roc}(b), we find that our pipeline is robust to variations in kernels by consistently achieving an AUC of $0.78$ on all cases.

\section{Relation to Existing Work}
\label{sec:related}
% \begin{itemize}
%     \item Why is PE detection challenging? explain with references
%     \item What is the pre-deep learning era approach to this? (Toboganning Algo from Nima, J.Lee ASU)
%     \item A brief timeline of how the methods have evolved - Segmentation/Object Detection and 2-stage pipelines (UNet, RCNN, YOLO, NoNewNet - uses 3D UNet, Dilated Dense UNet, weakly supervised instance segmentation (PRM), Unsupervised Landmark generation)
%     \item What is the latest and the greatest - Stanford PENet and where it stands.
%     \item Why do we need to design yet another method -- reduced time complexity, weak supervision, simplified architectures, interpretability. More importantly, generalization to all PE protocols (some of which have not been considered so far in benchmarking).
% \end{itemize}

In medical imaging applications, commonly deployed disease detection algorithms often involve multi-stage pipelines comprising both segmentation and classification models ~\cite{ardila2019end}. An early work on PE detection used custom feature extraction based on hierarchical anatomical segmentation of organs such as vessels, pulmonary artery, and aorta ~\cite{bouma2009automatic}. Though it appears natural to directly build a classifier model on the $3$D volumes, in practice, algorithms that first identify semantically meaningful candidate regions, and subsequently extract discriminative features from those regions to perform detection are found to be more effective. These methods are inspired by the success of such two-stage methods in object detection, examples include region-based RCNN ~\cite{girshick2015fast}. However, it is important to note that, adapting those techniques to problems in medical imaging have proven to be less trivial mainly for two reasons. One, these solutions require large datasets with ground truth in the form of bounding boxes that characterize regions of interest (ROIs) or dense annotations, which are usually harder to obtain in the clinical domain. Second, the models need to be capable of handling the heavy imbalance between number of positive cases against the more prevalent negative ROIs. Consequently, weakly supervised approaches have gained research interest. Methods that leverage information ranging from single-pixel labels ~\cite{anirudh2016lung} to approximate segmentation labels from class activation maps have been proposed~\cite{zhou2018weakly}. However, in the context of PE detection, most existing methods have relied exclusively on supervised learning with dense annotations, and the state-of-the-art solutions such as PENet~\cite{penet} utilize transfer learning from pre-trained models for effective detection.

% Also, improvements on weakly-supervised instance segmentation techniques have been made by leveraging class activation maps in an end-to-end trainable model ~\cite{Li_2018_CVPR, zhou2018weakly}. However, training end-to-end models to deal with large CT studies proves to be very compute intensive needing a large number of GPUs. Another typical formulation for problems under weak-supervision where instance-level labels are missing involves multiple instance learning (MIL). Consider a CT study to represent a bag of instances that only contains labels on the study-level as being positive or negative for a specific disease. In such a scenario, deep architectures that use recurrent neural networks in addition to CNNs called, ConvLSTMs, have been successfully applied to detect lung diseases such as emphysema \cite{braman2018disease}.

% Consequently, the latest results for PE detection was obtained by PENet \cite{penet}, a model which highlights the power of pre-training very deep architectures and using transfer learning for further fine-tuning with a large repository of densely annotated CTPA studies. Note, PENet comprises of a $77$-layer deep network pre-trained on $500$K video clips for $70$ hours achieving an AUROC of $0.79$ in retrospectively detecting PEs. 

 \section{Conclusion}
 \label{sec:conclusion}
 In this work, we present a generalizable two-stage pipeline for detecting pulmonary embolisms (PE) observed in $3$D CT images. The pipeline comprises of a context-augmented UNet model to generate segmentation masks, and a convolutional LSTM based classifier used in a MIL setting to detect PE. The proposed approach achieves state-of-the-art results on a challenging real-world dataset while alleviating need for dense annotations of CTs and using models with substantially lower number of parameters compared to prior art. We achieve AUC scores of $0.94$ on the validation set and $0.85$ on a test set of high-severity PEs. Further, our insights from the rigorous ablation studies conducted provide guidelines for designing effective disease detection pipelines.

%Our pipeline comprises of two stages, where \texttt{Stage $1$} uses a context-augmented UNet model to generate segmentation masks, and \texttt{Stage $2$} comprises of a convolutional LSTM based classifier used in a multiple instance learning (MIL) setting to detect PE. We evaluate the pipeline on a challenging real-world dataset containing studies from multiple hospitals, protocols, and imaging systems with various types of PE. Further, our insights from the rigorous ablation studies provides guidelines for designing effective disease detection pipelines.

\bibliographystyle{IEEEtran}
\bibliography{refs}

\end{document}